\documentclass{nature}

\usepackage{amssymb}
\usepackage{graphicx}
\usepackage[breaklinks,colorlinks=true,citecolor=blue,linkcolor=red,urlcolor=blue]{hyperref}
\usepackage{bm}
\usepackage{caption}
\usepackage[percent]{overpic}
\usepackage[symbol]{footmisc}

\makeatletter
\let\saved@includegraphics\includegraphics
\AtBeginDocument{\let\includegraphics\saved@includegraphics}
\renewenvironment*{figure}{\@float{figure}}{\end@float}
\makeatother

\newcommand{\ls}{\ell_\mathrm{s}}

\newcommand{\lbox}{L_\mathrm{box}}
\newcommand{\ldriv}{L_\mathrm{driv}}
\newcommand{\mach}{\mathcal{M}}
\newcommand{\rhocrit}{\rho_\mathrm{critical}}
\newcommand{\lj}{\ell_\mathrm{Jeans}}
\newcommand{\phix}{\phi_\mathrm{x}}
\newcommand{\phis}{\phi_\mathrm{s}}
\newcommand{\tturb}{t_\mathrm{turb}}

\newcommand{\cs}{c_\mathrm{s}}
\newcommand{\pc}{\mathrm{parsec}}

\newcommand{\suppl}{Supplementary Information}

\title{The sonic scale revealed by the world's largest supersonic turbulence simulation}

\author{Christoph Federrath$^{1}$, Ralf S.~Klessen$^{2,3}$, Luigi Iapichino$^{4}$, James R.~Beattie$^{1,5}$}

\begin{document}

\maketitle

\begin{affiliations}
\item Research School of Astronomy and Astrophysics, Australian National University, Canberra, ACT~2611, Australia
\item Universit\"{a}t Heidelberg, Zentrum f\"{u}r Astronomie, Institut f\"{u}r theoretische Astrophysik, Albert-Ueberle-Str.~2, 69120~Heidelberg, Germany
\item Universit\"at Heidelberg, Interdisziplin\"ares Zentrum f\"ur Wissenschaftliches Rechnen, Im Neuenheimer Feld 205, 69120~Heidelberg, Germany
\item Leibniz-Rechenzentrum der Bayerischen Akademie der Wissenschaften, Boltzmannstra{\ss}e~1, 85748~Garching, Germany
\item Science and Engineering Faculty, Queensland University of Technology, Brisbane, QLD~4000, Australia
\end{affiliations}

\begin{abstract}
Understanding the physics of turbulence is crucial for many applications, including weather, industry, and astrophysics.
In the interstellar medium (ISM)\cite{Ferriere2001,HennebelleFalgarone2012}, supersonic turbulence plays a crucial role in controlling the gas density and velocity structure, and ultimately the birth of stars\cite{MacLowKlessen2004,KrumholzMcKee2005,McKeeOstriker2007,HennebelleChabrier2008,Hopkins2013IMF,PadoanEtAl2014}. Here we present a simulation of interstellar turbulence with a grid resolution of $\mathbf{10048^3}$ cells that allows us to determine the position and width of the \emph{sonic scale} ($\ls$)---the transition from supersonic to subsonic turbulence. The simulation simultaneously resolves the supersonic and subsonic cascade, $v(\ell)\propto\ell^p$, where we measure $p_\mathrm{sup}=0.49\pm0.01$ and $p_\mathrm{sub}=0.39\pm0.02$, respectively. We find that $\ls$ agrees with the relation $\ls / L = \phis \mach^{-1/p_\mathrm{sup}}$, where $\mach$ is the three-dimensional Mach number, and $L$ is either the driving scale of turbulence or the diameter of a molecular cloud. If $L$ is the driving scale, we measure $\phis=0.42^{+0.12}_{-0.09}$, primarily because of the separation between the driving scale and the start of the supersonic cascade. For a supersonic cascade extending beyond the cloud scale, we get $\phis=0.91^{+0.25}_{-0.20}$. In both cases, $\phis\lesssim1$, because we find that the supersonic cascade transitions smoothly to the subsonic cascade over a factor of $\sim3$ in scale, instead of a sharp transition. Our measurements provide quantitative input for turbulence-regulated models of filament structure and star formation in molecular clouds.
\end{abstract}

\begin{figure}
\centerline{\includegraphics[width=0.9\linewidth]{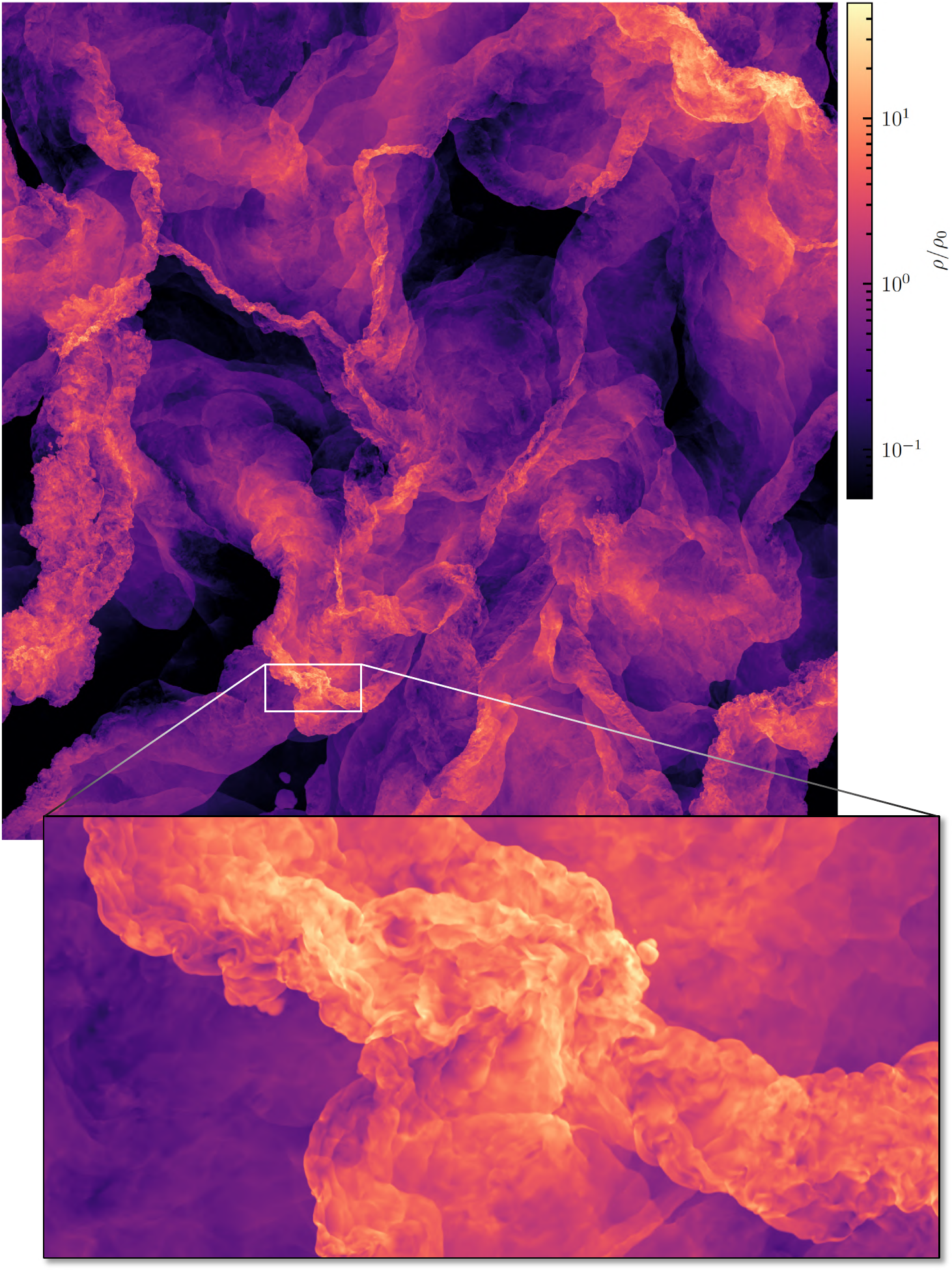}}
\caption{\label{fig:vis}
Slice through the gas density contrast (density $\rho$ divided by mean density $\rho_0$), showing the full simulation domain of the and a zoom-in onto a dense region with multiple shock interactions.}
\end{figure}

\noindent Here we present a simulation in which the sonic scale of turbulence is identified and resolved. A visualisation of this simulation is shown in Fig.~\ref{fig:vis}. A 3D animation is available online\footnote[1]{\scriptsize\url{https://www.mso.anu.edu.au/~chfeder/pubs/sonic_scale/Federrath_sonic_scale_lowres.mp4}.}. The theoretical prediction for the sonic scale is given by\cite{McKeeOstriker2007,FederrathKlessen2012,Hopkins2013IMF},
\begin{equation} \label{eq:ls}
\ls = \phis L (\sigma_v/\cs)^{-1/p},
\end{equation}
where $L$ is either the driving scale of the turbulence or the diameter of a molecular cloud. The ratio \mbox{$\sigma_v/\cs\equiv\mach$} is the three-dimensional (3D) turbulent Mach number\cite{MacLowKlessen2004,McKeeOstriker2007,PadoanEtAl2014}, i.e., the ratio of the volume-weighted 3D velocity dispersion ($\sigma_v$) on scale $L$ to the gas sound speed ($\cs$), with constant $\cs=$, however, we also analyse simulations with heating and cooling (Fig.~\ref{fig:mach_cool} in the Methods section). The exponent $p\approx0.5$ appears in the velocity dispersion -- size relation\cite{Larson1981}, $\sigma_v(\ell)\propto\ell^p$, as found in observations of molecular clouds\cite{OssenkopfMacLow2002,HeyerBrunt2004,RomanDuvalEtAl2011}. The factor $\phis$ encapsulates our lack of understanding of where exactly the sonic scale is located. Eq.~(\ref{eq:ls}) is a key ingredient for modern theories of star formation, allowing us to derive the star formation rate\cite{KrumholzMcKee2005,FederrathKlessen2012,PadoanEtAl2014}, the filament width in molecular clouds\cite{AndreEtAl2014,Federrath2016}, and the stellar initial mass function\cite{HennebelleChabrier2008,Hopkins2013IMF,OffnerEtAl2014}, all of which are outstanding problems in astrophysics. For these theories to have predictive power, it is therefore critical to determine $\ls$.

In order to measure $\ls$, we solve the equations of compressible hydrodynamics on a three-dimensional (3D), periodic grid with $10048^3$ grid cells, using a modified version of the FLASH code, which runs a factor of $\sim3.6$ faster than the public version (see Methods). The simulation was performed on 65536~compute cores, consumed approximately 50~million core hours and produced a total of 2~PB of data, split into 91~3D outputs, spanning 9~turbulent crossing times. The turbulent Mach number was set to $\mach(L)=4.1$ on the driving scale $L$, determined by the strength of the turbulence driving (we also test for varying $\mach(L)$ in Fig.~\ref{fig:mach_cool} in the Methods section). According to the prediction of Eq.~(\ref{eq:ls}), $\mach(L)=4.1$ would place $\ls$ at about 1/16th of $L$, allowing us to resolve $\ls$, and both the supersonic and subsonic cascades, respectively above and below $\ls$. This is despite the influence of dissipation, which sets in on scales of about 20--30 grid cells\cite{FederrathDuvalKlessenSchmidtMacLow2010}, i.e., for $\ell/L\lesssim30/10048$ (see Methods).

In order to construct $\mach(\ell)$ as a function of scale $\ell$, from which we can directly determine $\ls$, we compute the total (transverse plus longitudinal) 2nd-order velocity structure function\cite{KonstandinEtAl2012}, $\mathrm{SF_2}(\ell/L) = \langle |\mathbf{v}(\mathbf{r}) - \mathbf{v}(\mathbf{r+\bm{\ell}}/L)|^2 \rangle_\mathbf{r}$, where we only keep the dependence on the absolute value of the spatial separation ($\ell = |\bm{\ell}|$) between the gas velocity $\mathbf{v}(\mathbf{r})$ at position $\mathbf{r}$ and the velocity $\mathbf{v}(\mathbf{r+\bm{\ell}}/L)$ at position $\mathbf{r+\bm{\ell}}/L$. The angle bracket operation $\langle\dots\rangle_\mathbf{r}$ denotes the average over a sufficiently large sample of independent spatial positions $\mathbf{r}$, such that $\mathrm{SF_2}(\ell/L)$ is statistically converged (Fig.~\ref{fig:sfstats}) in the Methods section. The structure function was computed and averaged for a period of 7~turbulent crossing times (where one crossing time is defined as $\tturb=L/\sigma_v$), between \mbox{$2$--$9\,\tturb$}, i.e., over 71~3D snapshots of the $10048^3$ data (note that it takes $2\,\tturb$ to reach fully-developed turbulence\cite{FederrathDuvalKlessenSchmidtMacLow2010}).

\begin{figure}
\centerline{\includegraphics[width=0.99\linewidth]{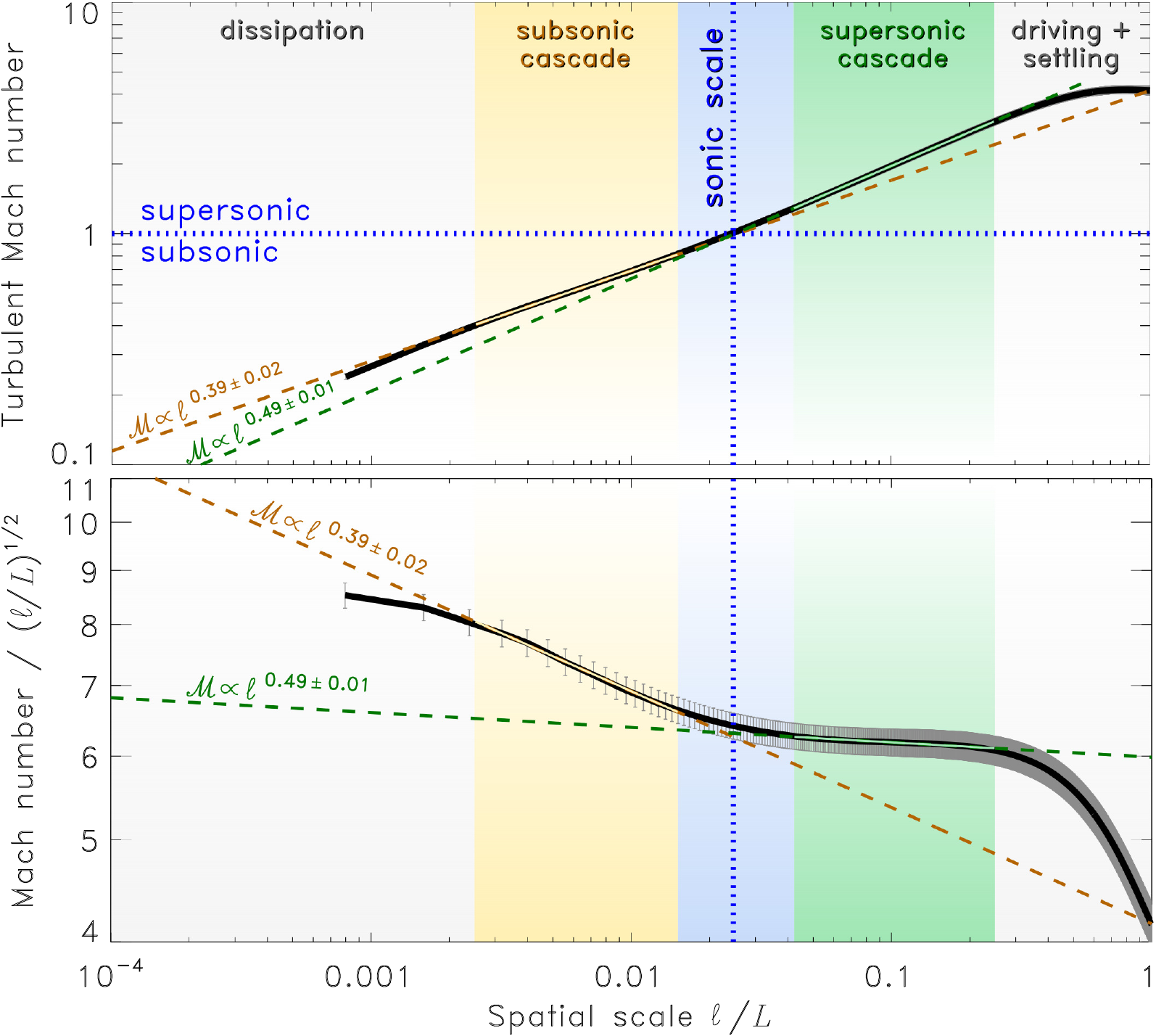}}
\caption{\label{fig:sf}
{\bf Top panel:} Turbulent Mach number ($\mach$) as a function of scale $\ell$ in units of the driving scale $L=\lbox/2$, where $\lbox$ is the side length of the computational box. $\mach(\ell/L)$ is computed from the 2nd-order structure function of our simulation with $10048^3$ grid cells. On the driving scale ($\ell=L$), we set the Mach number to $\mach=4.13\pm0.23$, shown by the flattening of function for $\ell/L\to1$. The 1$\sigma$ error bars were computed from the temporal fluctuations in the regime of full-developed turbulence. The sonic scale $\ls$ (vertical dotted line in the top and bottom panels) is defined where $\mach(\ls/L)=1$ (horizontal dotted line). The colour-shaded areas show different regimes of turbulence, as denoted at the top of top panel. {\bf Bottom panel:} Same as the top panel, but with $\mach$ compensated (divided) by $(\ell/L)^{1/2}$ to enhance the visibility of the change in slope across $\ls$. The dashed lines show power-law fits in the subsonic (gold) and supersonic (green) regimes of turbulence, with slopes of $0.39\pm0.02$ and $0.49\pm0.01$, respectively (in both the top and bottom panel). For the position and width of the sonic scale we find $\ls/L=0.025^{+0.007}_{-0.005}$, i.e., a full-width-half-maximum (FWHM) in $\ls$ corresponding to a factor of 3 in scale (blue shaded area).}
\end{figure}

Fig.~\ref{fig:sf} shows the time-averaged structure function (with error bars quantifying the 1$\sigma$ time fluctuations (the respective velocity power spectrum is shown in Fig.~\ref{fig:spect}) in the Methods section. In particular, we show the Mach number $\mach(\ell/L) = [\mathrm{SF_2}(\ell/L)/(2\cs^2)]^{1/2}$, constructed from $\mathrm{SF_2}$, where $\cs$ is the isothermal sound speed of the gas\cite{KonstandinEtAl2012}. The sonic scale is implicitly defined where \mbox{$\mach(\ls/L)=1$}. By determining on which scale the $\mach=1$, we find $\ls/L=0.025$ (vertical blue dotted line in Fig.~\ref{fig:sf}). Power-law fits with $\mach(\ell)\propto\ell^p$ in the subsonic (gold shaded area in Fig.~\ref{fig:sf}) and supersonic (green shaded area in Fig.~\ref{fig:sf}) regimes yield scaling exponents of $p_\mathrm{sub}=0.39\pm0.02$ and $p_\mathrm{sup} = 0.49\pm0.01$, respectively, consistent with the theoretical expectations in the two regimes of turbulence, i.e., $p_\mathrm{sub}\sim1/3$ for Kolmogorov turbulence\cite{Kolmogorov1941c} and $p_\mathrm{sup}\sim1/2$ for supersonic turbulence\cite{Burgers1948}. The subsonic slope is slightly steeper than the original Kolmogorov prediction of $p_\mathrm{sub}=1/3$, potentially because of some remaining effects from numerical dissipation (Fig.~\ref{fig:resolution-study} in the Methods section), but more likely because of the necessary intermittency\cite{FalgaronePetyHilyBlant2009} corrections for mildly compressible turbulence\cite{SchmidtFederrathKlessen2008}. We find that the transition region around the sonic scale (blue shaded area in Fig.~\ref{fig:sf}) is smooth and spans a full width half maximum (FWHM) factor of 3 in $\ell/L$, which corresponds to a Gaussian standard deviation of $\log_{10}(3/2.355)=0.105$. Thus, we measure $\log_{10}(\ls/L) = -1.608 \pm 0.105$ or $\ls/L=0.025^{+0.007}_{-0.005}$ for $\mach(L)=4.1$.

Assuming $\phis=1$ in Eq.~(\ref{eq:ls}), as done in previous works would suggest $\ls/L=\mach^{-2}=4.13^{-2}=0.059$ for the present simulation. This is different from our direct measurement of $\ls/L=0.025$, which implies $\phis=0.42^{+0.12}_{-0.09}$, i.e., the sonic scale is located on $1/\phis \sim 2.4$ times smaller scales. There are three reasons for this: \emph{i}) the supersonic cascade does not start right on the driving scale ($L$), but is located at $\ell/L\lesssim0.25$ (green shaded region in Fig.~\ref{fig:sf}; best seen in panel~b, where the compensated $\mathrm{SF_2}$ shows a significant change in slope at $\ell/L\sim0.25$), \emph{ii}) there is a smooth transition between the supersonic and subsonic cascade around the sonic scale, i.e., the power law implied by Eq.~(\ref{eq:ls}) does not extend all the way to the sonic scale, but smoothly flattens before that (best seen in Fig.~\ref{fig:sf} bottom panel), and \emph{iii}) the power-law exponent is not exactly $p=1/2$ in Eq.~(\ref{eq:ls}), although we find that it is very close to it ($p_\mathrm{sup}=0.49\pm0.01$; see Fig.~\ref{fig:sf}).

The $\phis=0.42$ result applies only, if $L$ is assumed to be the driving scale of the turbulence. However, the scale $L$ in Eq.~(\ref{eq:ls}) can also be interpreted as the molecular cloud scale (approximate diameter). Since the velocity dispersion -- size relation is usually observed to continue beyond the cloud scale\cite{OssenkopfMacLow2002,HeyerBrunt2004,RomanDuvalEtAl2011}, our results may suggest that the primary driving scale of ISM turbulence is on scales larger than the cloud itself. In that case, a flattening of the Mach number -- size relation would not occur around the cloud scale (although in the transition from $\ell\gtrsim L$ to $\ell\lesssim L$, we expect an increase in $\mach$, because of cooling; see Fig.~\ref{fig:mach_cool} in the Methods section). Thus, the supersonic scaling with $\sigma_v^{-2}$ would start immediately at the cloud scale, and we would find $\phis=0.91^{+0.25}_{-0.20}$. In that case, the shift between the driving scale and the start of the supersonic cascade is absent (point \emph{i} above), and primarily the fact that the sonic scale transitions smoothly from the supersonic to the subsonic cascade (point \emph{ii}) leads to $\phis\lesssim1$.

\begin{figure}
\centerline{\includegraphics[width=0.99\linewidth]{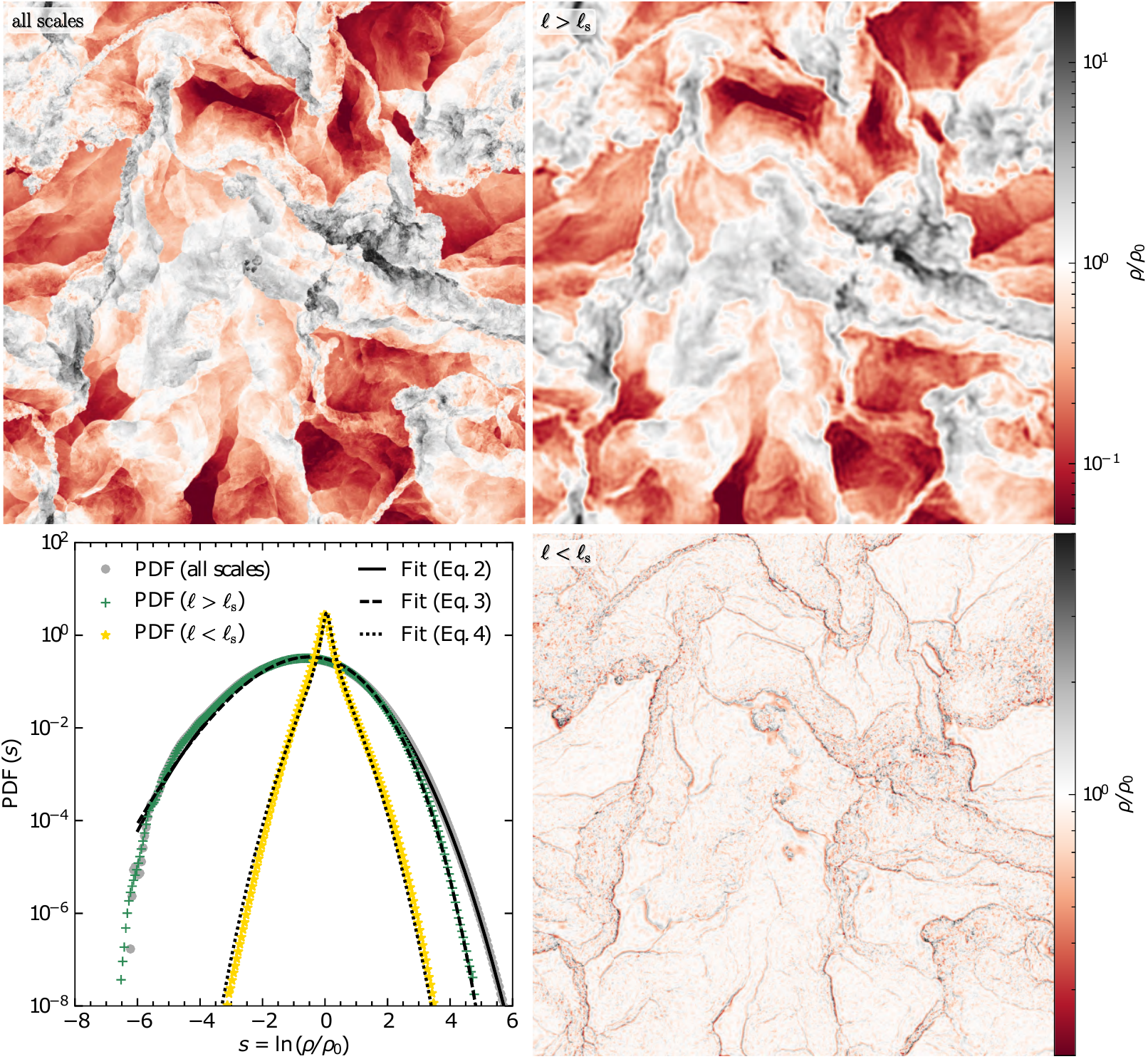}}
\caption{\label{fig:pdfs}
{\bf Top-left panel:} Slice through the full simulation density contrast $\rho/\rho_0$, including all scales, after 5~turbulent turnover times ($\rho_0$ is the mean density). {\bf Top-right panel:} Same as top-left panel, but the density field was Fourier-filtered such that only scales greater than the sonic scale ($\ell>\ls$) are retained. Note that top panels share the same colour map. {\bf Bottom-left panel:} Probability distribution functions (PDFs) of the logarithmic density contrast $s=\ln(\rho/\rho_0)$. The datapoints show the time-averaged density PDF data for the density field on all scales (silver circles), the Fourier-filtered density field keeping $\ell>\ls$ (green crosses), and $\ell<\ls$ (gold stars). 1$\sigma$ uncertainties (not shown for clarity) are of the order of twice the symbol size. Fits are shown as black lines; for the density field including all scales (solid; Eq.~\ref{eq:hopkins}), for $\ell>\ls$ (dashed; Eq.~\ref{eq:squire}), and for $\ell<\ls$ (dotted; Eq~\ref{eq:cauchy}). {\bf Bottom-right panel:} Same as the top-right panel, but Fourier-filtered to keep scales below the sonic scale ($\ell<\ls$).}
\end{figure}

Since the density structures around the sonic scale are believed to play a key role in star formation models\cite{KrumholzMcKee2005,HennebelleChabrier2008,FederrathKlessen2012,Hopkins2013IMF,PadoanEtAl2014}, we now study the probability distribution function (PDF) of the gas density on different scales\cite{SquireHopkins2017}. Fig.~\ref{fig:pdfs} (top-left panel) shows a slice through the density field after 5~turbulent turnover times. We see large density contrasts ranging over more than two orders of magnitude. Sharp edges between low-density and high-density regions are visible throughout the simulation domain, marking the position of strong hydrodynamic shock waves. The respective PDF of the full density field is shown as silver circles in Fig.~\ref{fig:pdfs} (bottom-left panel). For the density PDF covered by all scales, we fit the intermittency PDF model\cite{Hopkins2013PDF},
\begin{eqnarray} \label{eq:hopkins}
p(s) = I_1\left(2\sqrt{\lambda\,u}\right)\exp\left[-\left(\lambda+u\right)\right]\sqrt{\frac{\lambda}{\theta^2\,u}}\,,\nonumber\\
u\equiv\lambda/(1+ \theta)-s/ \theta \;\; (u\geq0), \quad  \lambda\equiv\sigma_s^2/(2 \theta^2)
\end{eqnarray}
where $I_1(x)$ is the 1st-order modified Bessel function of the first kind. Eq.~(\ref{eq:hopkins}) contains two parameters: the standard deviation of logarithmic density fluctuations $\sigma_s$, where $s=\ln(\rho/\rho_0)$, and an intermittency parameter $ \theta$. In the zero-intermittency limit ($\theta\to0$), Eq.~(\ref{eq:hopkins}) simplifies to a log-normal PDF, as commonly used in studies of the density statistics of supersonic turbulence in molecular clouds. We find $\sigma_s=1.21\pm0.12$ and $\theta=0.042\pm0.005$. The relatively low level of intermittency quantified by the small value of $\theta$ indicates that the PDF is nearly log-normal\cite{Hopkins2013PDF,Federrath2013}. The log-normal density dispersion, $\sigma_s=[\ln(1+b^2\mach^2)]^{1/2}\sim1.14$, with the turbulence driving parameter ($b=0.4$)\cite{FederrathDuvalKlessenSchmidtMacLow2010} used here matches the value of $\sigma_s$ measured from the density PDF reasonably well. Thus, for the turbulence including all scales, the density field is approximately log-normal, a central assumption in most star-formation models\cite{KrumholzMcKee2005,HennebelleChabrier2008,FederrathKlessen2012,Hopkins2013IMF,PadoanEtAl2014}.

After Fourier-filtering the density field and keeping only scales above the sonic scale ($\ell>\ls$), shown in Fig.~\ref{fig:pdfs} (top-right panel), we find a density distribution with large-scale structure similar to the full density field, but with small-scale fluctuations removed. As a result, the highest-density regions (which include the filamentary structures on and below the sonic scale) are smoothed to lower density. The PDF for densities retaining scales $\ell>\ls$ (shown as green crosses in panel~c) is well fit by a scale-dependent intermittency PDF model\cite{SquireHopkins2017} (we note a typo in Eq.~(13) in ref.\cite{SquireHopkins2017}, which misses a factor $1/T$; here $\theta\equiv T$; on the right-hand side in their definition of $u$),
\begin{eqnarray} \label{eq:squire}
p(s) = I_1\left(2\sqrt{\lambda\,u}\right)\exp\left[-\left(\lambda+u\right)\right]\sqrt{\frac{\lambda}{\theta^2\,u}}\,,\nonumber\\
u\equiv\lambda/(1+ \theta)-s/ \theta \;\; (u\geq0), \quad  \lambda\equiv\xi(1+1/\theta)\ln(L/\ell).
\end{eqnarray}
Note that this equation is identical to Eq.~(\ref{eq:hopkins}), with the exception of the definition of $\lambda$, which (instead of $\sigma_s$) now contains two additional parameters, namely $\xi$ and $L/\ell$. The parameter $L/\ell$ describes the scale dependence of the density PDF and we fix it to $L/\ell=L/\ls=40.5$, directly based on our measurement of $\ls/L=0.025$ above. The fitted parameter $\theta=0.090\pm0.010$ indicates somewhat higher intermittency compared to the full density field, probably because the smoother, less-intermittent subsonic scales had been filtered out and the density PDF is more skewed. The fit parameter $\xi=1.60\pm0.20$ is related to the fractal dimension $D=3-\xi=1.4\pm0.2$ of the shocks\cite{SquireHopkins2017}, i.e., somewhere between pure lines ($D=1$) and pure sheets ($D=2$), broadly consistent with the typical 3D fractal dimension $D+1=2.4\pm0.2$ of molecular clouds\cite{Scalo1990,SanchezEtAl2005,FederrathKlessenSchmidt2009,BeattieEtAl2019a}.

Finally, Fig.~\ref{fig:pdfs} (bottom-right panel) shows the Fourier-filtered density field, but this time keeping only scales below the sonic scale ($\ell<\ls$). We see the filamentary shock structures stand out, which separate low-density pre-shock and high-density post-shock regions. Thus, these shock structures occurring on the sonic scale mark the transition from the supersonic cascade (outside the filamentary shocks) to the subsonic cascade (inside the filamentary shocks). The respective density PDF (gold stars in panel~c) is much narrower, peaks at about $s=0$, and has extended power-law wings around the peak, one of the key signatures of intermittent structures\cite{SchmidtFederrathKlessen2008,Hopkins2013PDF}. In order to model this behaviour, we fit a Cauchy-Lorentz distribution multiplied by a normal distribution in $s$,
\begin{equation} \label{eq:cauchy}
p(s) = \frac{p_0}{\pi w \left[1 + \left(\frac{s-s_0}{w}\right)^2\right]} \times \frac{1}{(2\pi\sigma_0^2)^{1/2}}\exp\left[-\frac{1}{2}\left(\frac{s-s_0}{\sigma_0}\right)^2\right],
\end{equation}
and find the fit parameters $p_0=1.80\pm0.34$, $s_0=0.05\pm0.01$, $w=0.11\pm0.02$, and $\sigma_0=0.66\pm0.07$. This function fits the PDF data very well, with the Cauchy-Lorentz factor describing the extended power-law wings (intermittency) of the distribution. The total standard deviation of the PDF is $\sigma_s=0.25\pm0.01$ in the log-density $s$ or $\sigma_\rho/\rho_0=0.30\pm0.01$ directly in density contrast $\rho/\rho_0$, for $\ell<\ls$, consistent with the relation $\sigma_\rho/\rho_0=b\mach$ [ref.\cite{PadoanNordlundJones1997,FederrathDuvalKlessenSchmidtMacLow2010}] for $\mach=1$ at the sonic scale and $b\sim0.3$, indicating that the turbulence around and below the sonic scale is primarily solenoidal\cite{FederrathDuvalKlessenSchmidtMacLow2010}, as would be expected for subsonic turbulence. For real star-forming molecular clouds, it is in these subsonic, filamentary regions dominated by solenoidal turbulence where gravity is able to overcome the turbulent energies, and local collapse of the gas leads to star formation. Thus, we conclude that the sonic scale may serve as a key ingredient for setting the width of molecular-cloud filaments (Fig.~\ref{fig:filamentpdf} in the Supplementary Information section) and for controlling the critical density of star formation (see {\suppl} section).

\begin{methods}

Here we describe the basic numerical setup and highlight some of the modifications necessary to enable us to perform a simulation with the unprecedented resolution of $10048^3$ computational elements. The simulations were run on SuperMUC Phase~1 at the Leibniz Supercomputing Centre (LRZ)\footnote[7]{\scriptsize\url{https://doku.lrz.de/display/PUBLIC/Decommissioned+SuperMUC}}. The system consists of 9216~nodes, each containing two 8-core Intel\textsuperscript{\textregistered} Xeon\textsuperscript{\textregistered}\ E5-2680 processors (code-named Sandy Bridge) with a clock frequency of $2.7\,\mathrm{GHz}$. This work is based on a simulation using 4096 such nodes, for a total of 65536~compute cores and 50~million core hours.

\subsection{Basic numerical code\\}
We employ a modified version of the hydrodynamical code FLASH\cite{FryxellEtAl2000,DubeyEtAl2008}, based on the public release version~4. Our code uses the state-of-the-art, positivity-preserving MUSCL-Hancock HLL5R Riemann scheme\cite{WaaganFederrathKlingenberg2011} to solve the compressible Euler equations of hydrodynamics in three dimensions:
\begin{equation} \label{eq:hydro}
\def\arraystretch{1.1}
\begin{array}{@{}l@{}}
\partial_t \rho + \nabla\cdot\left(\rho \vec{v}\right)=0 ,\\
\partial_t\!\left(\rho \vec{v}\right) + \nabla\cdot\left(\rho\vec{v}\!\otimes\!\vec{v}\right) + \nabla p = \rho{\vec{F}}, \\
p = \rho\cs^2,
\end{array}
\end{equation}
where $\rho$ and $\vec{v}$ are the gas density and the three velocity components, respectively, and $\otimes$ denotes the dyadic product. The last equation is the equation of state of isothermal gas, relating the gas pressure $p$ and the density by the constant sound speed $\cs$, replacing the energy equation in case of isothermal gases (while the main simulation is isothermal, we also explore simulations with heating and cooling, where the energy equation is solved; see Fig.~\ref{fig:mach_cool}). In order to drive turbulence, a specific (per unit mass) forcing term $\vec{F}$ is applied in the momentum equation (details below). This set of equations including the forcing term is the standard approach in modelling supersonic ISM turbulence\cite{PorterPouquetWoodward1992ThCFD,KritsukEtAl2007,SchmidtEtAl2009,Federrath2013}.

\subsection{Turbulence driving\\}
To drive turbulence with a given large-scale Mach number of $\mach(L)$, we apply a stochastic forcing term ${\vec{F}}$ in Eq.~(\ref{eq:hydro})\cite{EswaranPope1988,SchmidtEtAl2009,FederrathDuvalKlessenSchmidtMacLow2010}. The forcing is constructed in Fourier space such that kinetic energy is injected at the smallest wave numbers \mbox{($1<k\lbox/2\pi<3$)} or equivalently, on the largest scales \mbox{($1/3 < \ell/\lbox < 1$)}, where $\lbox=2L$, is the computational box size. The power spectrum of the driving amplitude follows a parabolic function in wavenumber $k=(k_x^2+k_y^2+k_z^2)^{1/2}$, such that the peak of the parabola is located at $k=k_\mathrm{driv}=2\times(2\pi/\lbox)=2\pi/L$, and the amplitude falls off on both sides of the peak to reach identically zero at exactly $k=1\times(2\pi/\lbox)$ and $k=3\times(2\pi/\lbox)$. Thus, the main driving scale is defined as $L\equiv\ldriv=\lbox/2$ throughout this study. Using this parabolic spectrum for the driving within \mbox{$1<k\lbox/2\pi<3$} restricts the driving to a narrow range with the peak power injected at $L=\lbox/2$, allowing for a self-consistent development of the turbulence on smaller scales, as routinely done in systematic turbulence studies\cite{HaugenBrandenburgMee2004,SchekochihinEtAl2007}. We decompose the forcing into its solenoidal (divergence-free) part and its compressive (curl-free) part and use the natural mixture of modes to drive the turbulence\cite{FederrathKlessenSchmidt2008,FederrathDuvalKlessenSchmidtMacLow2010,Federrath2013}. In the context of turbulence driving, we note that the supersonic scaling range (marked in Fig.~\ref{fig:sf} and Fig.~\ref{fig:spect} below) does not start immediately at $\ell/\lbox\leq1/3$ (where the driving range ends), but roughly at $\ell/\lbox\leq 1/8$ or equivalently $\ell/L\leq 1/4$. The reason for this is that it takes some separation in scale between the driving scale and the start of the supersonic scaling range for the turbulence to reach a fully-developed, self-similar state. Thus, the scales $1/4\leq\ell/L\leq1$ cover a `driving~+~settling' range (see Fig.~\ref{fig:sf} and Fig.~\ref{fig:spect} below). It is a general property of turbulence that the self-similar turbulent cascade can only start on scales somewhat separated from the driving scale\cite{Frisch1995,Pope2000}, which is part of the reason why $\phis<1$ in Eq.~(\ref{eq:ls}), as determined in the main part of the study, with the other reason being that the transition to the sonic scale is smooth and not abrupt, but instead occurs over factor of 3 in scale (Fig.~\ref{fig:sf}).

\subsection{Code optimisation\\}
Reaching the required numerical resolution of $10048^3$ cells could only be achieved with extreme measures to optimise the performance of the code and to minimise the memory requirements. Specifically, we have modified the standard FLASH version in the following three ways: First, we reduced the number of stored 3D fields to the bare minimum required for these simulations. For instance, the FLASH code typically stores the adiabatic index, the pressure, the internal and the total energy in individual 3D fields, but we do not need them here (only the gas density and the three velocity components need to be stored). Second, we removed all calls to the equation-of-state routines in the FLASH code such that these operations are performed inline, directly in the Riemann solver. Third, we built a hybrid code that stores all gas variables in single precision (4~bytes per floating-point number), but performs critical operations in double-precision arithmetic (8~bytes per floating-point number), retaining the accuracy of the previous full double-precision computations, but reducing the memory footprint by a factor of~2. Any operation that requires double precision is implemented by explicitly forcing the code to perform such critical operations in double precision, such as solving the hydrodynamical equations in the Riemann solver and computing global integral flow quantities for high-resolution time monitoring of the simulation. These combined efforts have significantly reduced the computational time and the required amount of Message-Passing-Interface (MPI) communication, as well as the overall memory consumption, such that the code is now 3.6 times faster and requires 4.1 times less memory for our particular problem setup, while retaining full accuracy with our new hybrid-precision scheme (see Fig.~\ref{fig:hb}).

\begin{figure}
\centerline{\includegraphics[width=0.99\linewidth]{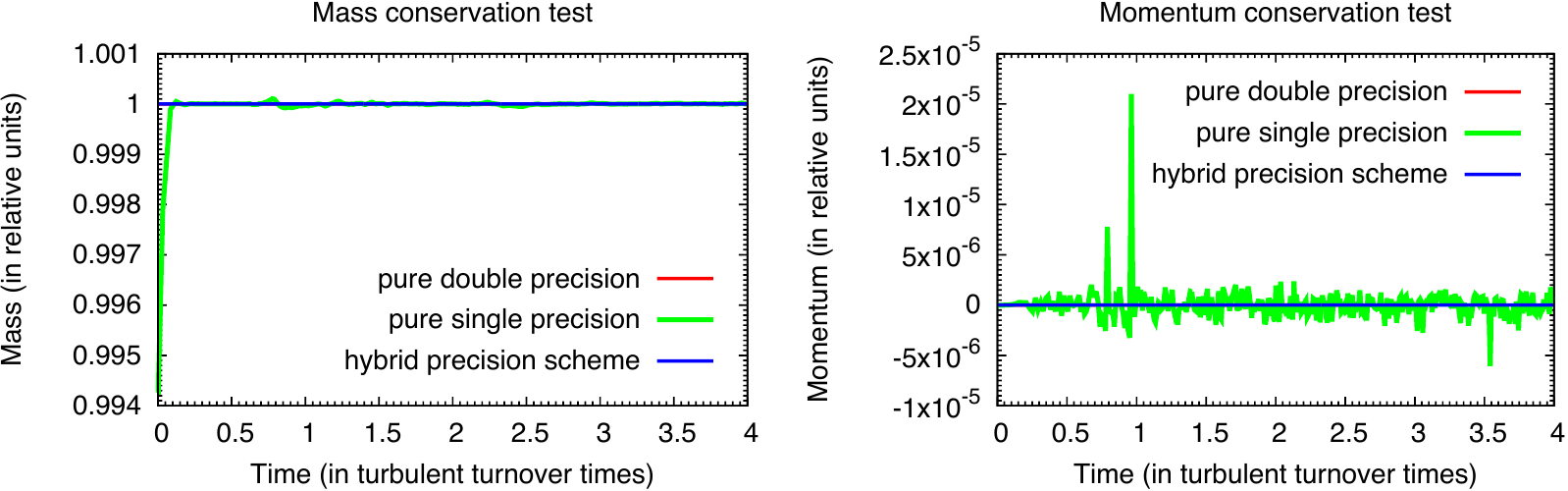}}
\caption{\label{fig:hb}
{\bf Left Panel:} Comparison of the pure double-precision and pure single-precision schemes with our new hybrid-precision scheme for modelling supersonic and subsonic turbulence. The gas mass as a function of time is well conserved in our hybrid-precision scheme, shown as the straight blue line (identical to the pure double-precision scheme: red line; not visible, because it lies exactly behind the blue line), while significant errors arise in the pure single-precision scheme (green line). {\bf Right panel:} Same as left panel, but for the gas momentum.}
\end{figure}

\subsection{Data structure and domain decomposition\\}
FLASH uses a block-structured parallelisation scheme. For the main production simulation, each three-dimensional computational block is distributed onto one single compute core. Each block contains $157\times314\times314$ cells, resulting in ($157\times64$, $314\times32$, $314\times32$) $=$ ($10048$, $10048$, $10048$) cells in each spatial direction, which we distribute over a total of $64\times32\times32=65536$~cores.

\subsection{File I/O\\}
FLASH is parallelised with MPI. File I/O is based on the hierarchical data format, version~5 (HDF5) library. Since our production run uses 65536~compute cores and produces 91~output files with about $20\,\mathrm{TB}$ each, efficient file I/O is extremely important. In order to achieve the highest efficiency when reading and writing the files, we use MPI-parallel HDF5 together with a split-file approach. This means that each core writes simultaneously to the filesystem, grouping data from 256~cores together into a total of 256~files per output dump. This proved to be the most efficient way and provides an I/O throughput that is close to the physical maximum of about $160\,\mathrm{GB/s}$ achievable on the SuperMUC General Parallel File System (GPFS).

\subsection{Dissipation and numerical resolution requirements\\}
An important caveat in all turbulence simulations is the numerical resolution. In the study presented here, the dissipation is purely numerical, which minimises the amount of dissipation and results in a maximum possible scaling range. However, this also means that the dissipation scale depends on the numerical resolution, which shifts to smaller and smaller scales as the resolution is increased. In order to find the converged, physical scales, we need to perform resolution studies. The approach of relying on numerical dissipation furthermore needs to be tested against simulations that use explicit viscosity terms. Typically, one finds good agreement\cite{BenziEtAl2008,FederrathEtAl2011}, with scales larger than 20--30 grid cells not significantly affected by numerical dissipation\cite{FederrathDuvalKlessenSchmidtMacLow2010,FederrathSurSchleicherBanerjeeKlessen2011}. A numerical resolution study is provided in Fig.~\ref{fig:resolution-study}. This compares our main simulation with $10048^3$ grid points (solid line), with versions of the same simulation, but computed at $(2\times)^3$ and $(4\times)^3$ lower resolution, shown as the dotted and dashed lines, respectively. We find sonic scales, of $\ls/L=0.0247$, $0.0231$, and $0.0199$, for resolutions of $10048^3$, $5024^3$, and $2512^3$, respectively. This corresponds to relative differences of $\ls$ in the $10048^3$ simulation of 6.5\% and 19.4\% compared to the $5024^3$ and $2512^3$ simulation, respectively, or a relative difference of $6.5\%$ between $10048^3$ and $5024^3$, and a relative difference of $13.9\%$ between $5024^3$ and $2512^3$. Estimating the convergence behaviour based on a geometric series where the difference in sonic scale decreases with a factor of $6.5\%/13.9\%=0.47$ every time the linear resolution is doubled, means that the sonic scale measured here is converged to within 6\% uncertainty of the infinite-resolution limit. This limit would suggest an estimate of $\ls/L=1.06\times0.0247=0.0262$, however, since this is well within the scale range covered by the transition from supersonic to subsonic turbulence (c.f.~Fig.~\ref{fig:sf}), we will simply take the value of $\ls=0.025$ directly measured in the $10048^3$ simulation for further analyses. Fig.~\ref{fig:resolution-study} also reveals that a resolution of $10048^3$ cells is required to accurately measure the scaling exponent on scales smaller than the sonic scale, as the dissipation scale moves a factor of 2 and 4 closer to the sonic scale in the $5024^3$ and $2512^3$ simulation, respectively, cutting too much into the subsonic cascade. Thus, we find that for fully recovering the scaling slope in both the supersonic and subsonic regimes requires numerical resolutions of $\gtrsim10000^3$ grid cells.

\begin{figure}
\centerline{\includegraphics[width=0.9\linewidth]{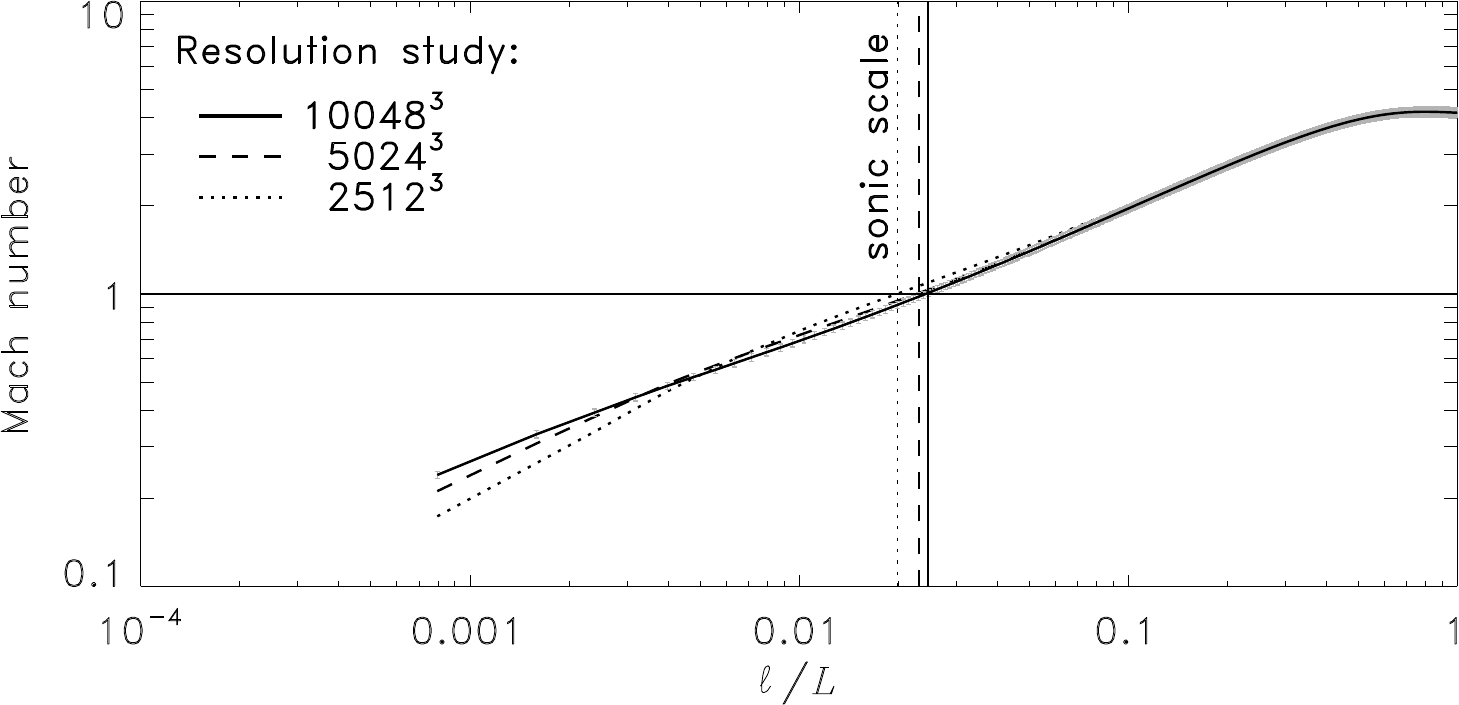}}
\caption{\label{fig:resolution-study}
Same as Fig.~\ref{fig:sf} (top panel), but for numerical grid resolutions of $5024^3$ cells (dashed) and $2512^3$ cells (dotted), in addition to our main run with $10048^3$ cells (solid line with 1$\sigma$ uncertainties shown in grey; note that the uncertainties on the dashed and dotted lines are similar to the ones shown on the solid line, but are omitted for clarity). The position of the sonic scale is converged to within 6\% of the extrapolated infinite-resolution limit. Measuring the width of the sonic transition range and the scaling in the subsonic regime requires $\gtrsim10048^3$ cells, as can be seen by the deviations below $\ls$ in the $5024^3$ and $2512^3$ resolution cases.
}
\end{figure}

\subsection{Velocity power spectrum\\}
A standard diagnostic in turbulence analyses is the power spectrum, which serves as a check of the method and the basic results. Fig.~\ref{fig:spect} shows the velocity power spectrum of our $10048^3$ simulation. This contains essentially the same information as the structure function shown in Fig.~\ref{fig:sf} (top panel), but instead of presenting the velocity fluctuations as a function of scale $\ell/L$, the power spectrum shows them as a function of wave number $k=2\pi/\ell$ in units of the driving wave number $K=2\pi/L$, i.e., in a Fourier-space representation instead of the real-space representation of Fig.~\ref{fig:sf}. The power spectrum is locally affected by the so-called `bottleneck effect'\cite{Falkovich1994,DoblerEtAl2003,VermaDonzis2007,SchmidtHillebrandtNiemeyer2006} close to the dissipation range, which causes a piling-up of kinetic energy that can affect the power-law scaling in the subsonic scaling range. The bottleneck effect has been observed in turbulence experiments\cite{Frisch1995}, and we also observe this bottleneck effect here in our simulations. The bottleneck effect has some consequences for the accuracy of the power-law fit in the subsonic scaling range, where we find $P_v \propto k^{-1.76\pm0.04}$ through fitting in Fig.~\ref{fig:spect}. Given the definition of the velocity power spectrum\cite{Frisch1995}, $P_v \propto d(v^2)/dk$, this corresponds to a velocity (or Mach number) scaling of $v\propto\ell^{-(-1.76+1)/2}\propto\ell^{0.38}$, which, despite the bottleneck effect, is consistent with the corresponding structure function fit in the subsonic scaling range in Fig.~\ref{fig:sf}, where we found $v \propto \ell^{0.39\pm0.02}$. The same holds for the supersonic scaling range, which is not affected by the bottleneck at all, where we find $P_v \propto k^{-1.99\pm0.02}$, equivalent to $v\propto\ell^{0.50}$, in very good agreement with the direct structure function fit of $v\propto\ell^{0.49\pm0.01}$ in the supersonic scaling range in Fig.~\ref{fig:sf}.

\begin{figure}
\centerline{\includegraphics[width=0.99\linewidth]{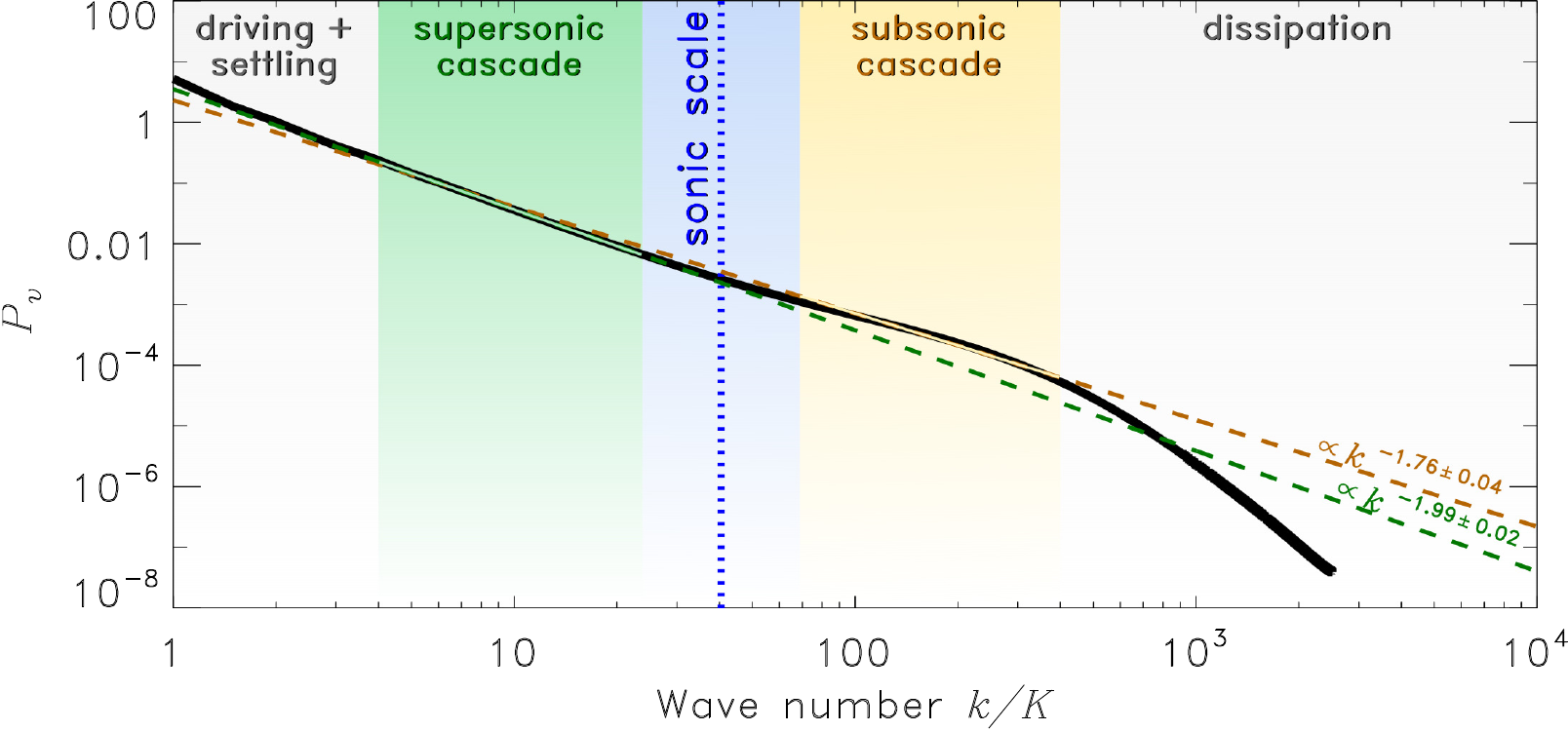}}
\caption{\label{fig:spect}
Same Fig.~\ref{fig:sf} (top panel), but showing the velocity power spectrum $P_v$, as a function of wave number $k$ normalised to the driving wave number $K=2\pi/L$. The power spectrum contains the same basic information as the structure function shown in Fig.~\ref{fig:sf}. We find $P_v \propto k^{-1.99\pm0.02}$ for the supersonic cascade and $P_v \propto k^{-1.76\pm0.04}$ for the subsonic cascade, with the sonic scale in between, consistent with the results derived from Fig.~\ref{fig:sf}.}
\end{figure}

\subsection{Statistical convergence of the structure functions\\}
Computing structure functions of turbulence (such as the ones shown in Fig.~\ref{fig:sf} and Fig.~\ref{fig:mach_cool}) is extremely challenging as it requires a very large number of sampling pairs to accurately converge on the intrinsic statistical properties of the turbulent flow. In order to achieve convergence on all scales measured by the structure function, one needs to be careful to distribute the workload and number of statistical samples equally across all scales involved. This is non-trivial, because the full computation would involve all available flow points. For example, for a simulation of $N^3$ cells, this would require operations on $(N^3)^2=N^6$ points, which for any reasonably large $N$ (e.g., our required numerical resolution of $N\sim10000$; see Fig.~\ref{fig:resolution-study}), would result in such a large numerical problem size that it would not be computable on even the largest supercomputers available to date (e.g., for $N=10^4$, the full requirement would be to perform operations on $10^{24}$ points). Therefore, one must choose a statistically representative subset of all available flow points to sample the structure functions with sufficient accuracy. To find a statistically converged solution of the structure function, we carry out a sampling study, shown in Fig.~\ref{fig:sfstats}. We compare the structure function computed based on different numbers of sampling points: $2\times10^8$, $2\times10^9$, $2\times10^{10}$, $2\times10^{11}$, and $2\times10^{12}$ pairs. This demonstrates statistical convergence of the 2nd-order structure function on all relevant scales for a sample size of $\gtrsim10^{12}$ pairs per time snapshot. For the final result shown in Fig.~\ref{fig:sf}, a total of 71~snapshots were used, each with $2\times10^{12}$ sampling pairs, resulting in a total of $\sim10^{14}$ sampling pairs contributing to the structure function shown in Fig.~\ref{fig:sf}.

\begin{figure}
\centerline{\includegraphics[width=0.9\linewidth]{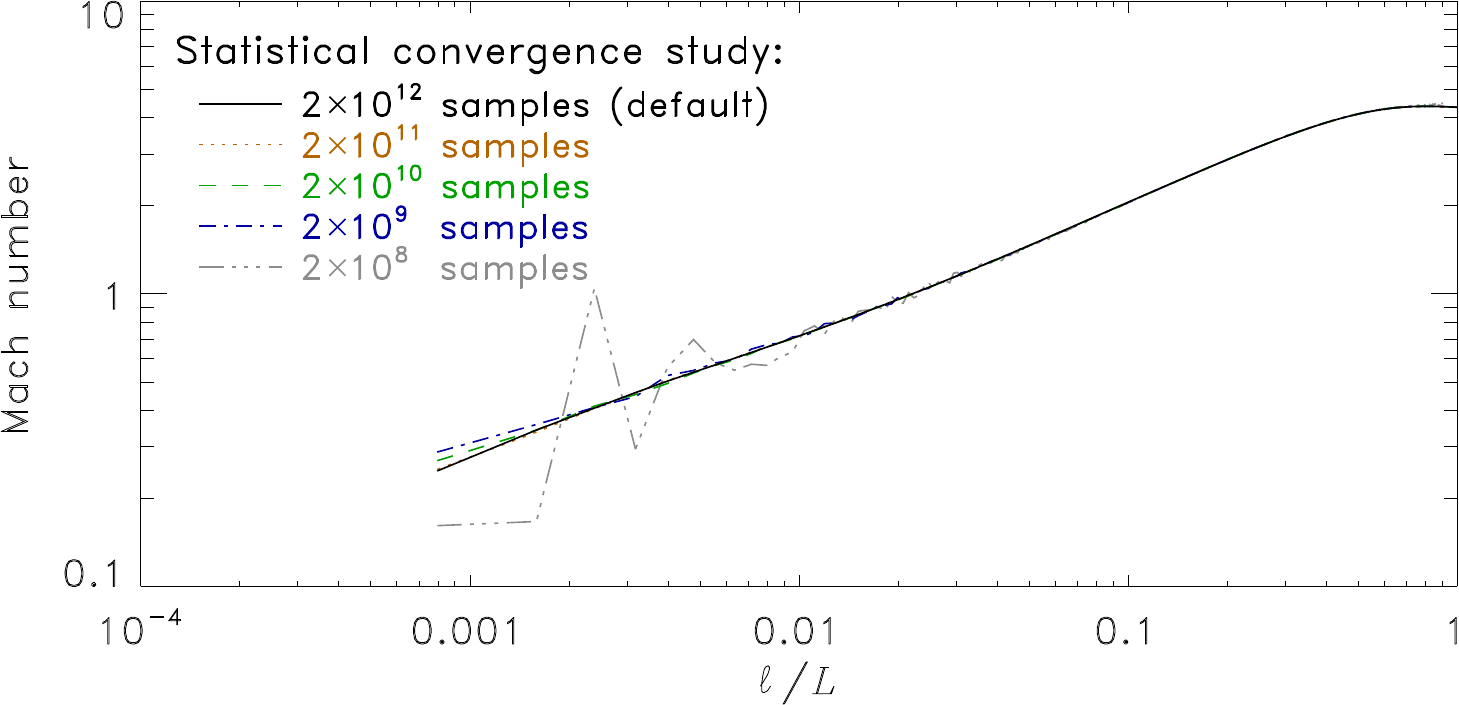}}
\caption{\label{fig:sfstats}
Same as Fig.~\ref{fig:sf} (top panel), but for a single time snapshot (at 5~turbulent crossing times), and the structure functions shown here were computed with different numbers of sampling points: $2\times10^8$, $2\times10^9$, $2\times10^{10}$, $2\times10^{11}$, and $2\times10^{12}$ pairs. This demonstrates statistical convergence of the 2nd-order structure function on all relevant scales for a sample size of $\gtrsim10^{12}$ pairs per time snapshot.}
\end{figure}

\subsection{Influence of the choice of Mach number and thermodynamical modelling\\}
In order to study the effects of changing the sonic Mach number on the driving scale $L$, and relaxing the isothermal approximation used in the main simulation, we run an additional set of simulations that varies the Mach number from $\sim2$ to $8$, and another set of simulations that uses cooling to control the gas temperature instead of fixing the temperature. The cooling rate follows a density--temperature curve developed based on the work by Koyama~\&~Inutsuka~(2002)\cite{KoyamaInutsuka2002} and V\'{a}zquez-Semadeni~et~al.~(2007)\cite{VazquezSemadeniEtAl2007}, and used in previous studies\cite{KoertgenFederrathBanerjee2019,MandalFederrathKoertgen2020}.

\begin{figure}
\centerline{\includegraphics[width=0.9\linewidth]{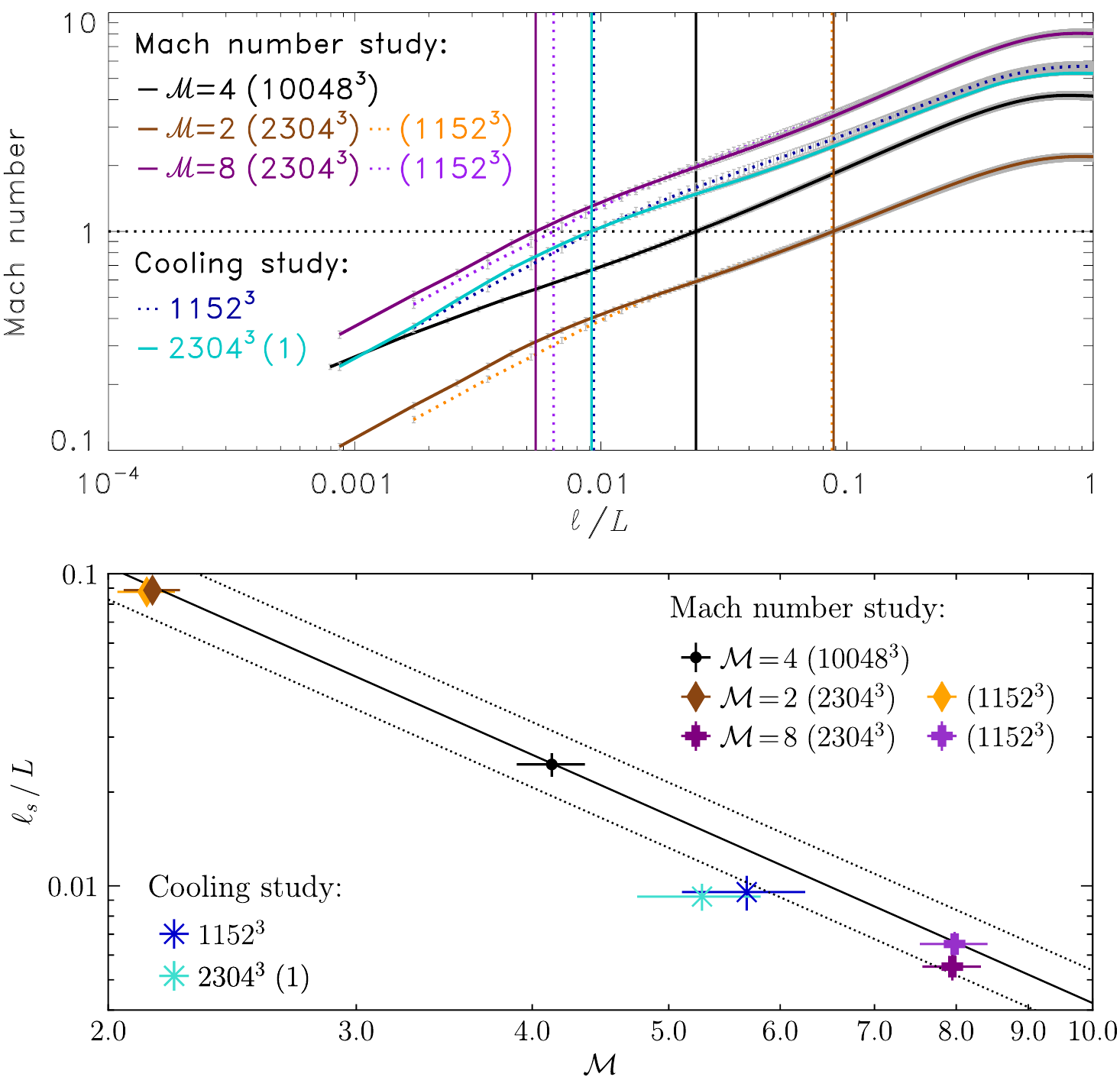}}
\caption{\label{fig:mach_cool}
{\bf Top panel:} Same as Fig.~\ref{fig:sf} (top panel), but for models with different Mach number (`Mach number study') and without the isothermal approximation (`Cooling study'), at different numerical resolutions: 
`\mbox{$\mach\!=\!4$ ($10048^3$)}' (black), `\mbox{$\mach\!=\!2$ ($2304^3$)}' (brown), `\mbox{$\mach\!=\!2$ ($1152^3$)}' (orange), `\mbox{$\mach\!=\!8$ ($2304^3$)}' (purple), `\mbox{$\mach\!=\!8$ ($1152^3$)}' (violet), `\mbox{Cooling $1152^3$}' (blue), and `\mbox{Cooling $2304^3\,(1)$}' (turquoise).
The vertical lines mark the location of the sonic scale in each of the different models, i.e., where the scale-dependent Mach number equals unity (horizontal dotted line). {\bf Bottom panel:} The sonic scale measured from the same models in the top panel, as a function of the Mach number ($\mach$) on the driving scale $L$. The solid and dotted lines show Eq.~(\ref{eq:ls}) with $p=1/2$ and $\phis=0.42^{+0.12}_{-0.09}$, as measured from the high-resolution ($10048^3$ grid cells) simulation in the main part of the study.}
\end{figure}

Fig.~\ref{fig:mach_cool} shows the results of this parameter study. Simulations `\mbox{$\mach\!=\!2$ ($2304^3$)}', `\mbox{$\mach\!=\!2$ ($1152^3$)}', `\mbox{$\mach\!=\!8$ ($2304^3$)}', and `\mbox{$\mach\!=\!8$ ($1152^3$)}', are identical to the main simulation, `\mbox{$\mach\!=\!4$ ($10048^3$)}', except that the turbulence driving was adjusted to produce Mach numbers of $\mach\sim2$ and $8$, respectively, and the grid resolutions were set to $1152^3$ and $2304^3$ grid cells, respectively, as indicated by the model label. The simulations labelled `\mbox{Cooling $1152^3$}' and `\mbox{Cooling $2304^3\,(1)$}' employ cooling as opposed to all the other simulations, which use an isothermal equation of state. The two cooling simulations use the same methods for heating and cooling of the gas described in Mandal et al.~(2020)\cite{MandalFederrathKoertgen2020}, but here we use a computational domain size of $20\,\pc$ (and hence a driving scale of $L=10\,\pc$), a velocity dispersion of $3\,\mathrm{km\,s^{-1}}$ and a mean gas number density of $100\,\mathrm{cm}^{-3}$ with a mean molecular weight of $2$, such that we are approximating the physical conditions for the transition from an atomic cold neutral medium with temperature $\sim100\,\mathrm{K}$ to a very cold molecular medium with temperature $\sim10\,\mathrm{K}$, similar to chemo-dynamical simulations\cite{GloverFederrathMacLowKlessen2010}, but here at resolutions of $1152^3$ and $2304^3$ grid cells. While these resolutions are enough to reach into the sonic scale, expected to occur at around $0.1\,\pc$ [ref.\cite{AndreEtAl2014,ArzoumanianEtAl2011,Federrath2016,ArzoumanianEtAl2018}], it is not sufficient for complete convergence (c.f., Fig.~\ref{fig:resolution-study}). However, this is a reasonable compromise given the high cost of these simulations, because they cannot make use of the hybrid-precision scheme. Thus, we could only evolve \mbox{Cooling $2304^3\,(1)$} for a single fully-developed turnover time for averaging in the interval $2\leq t/\tturb \leq 3$; all other simulations here were averaged over $2\leq t/\tturb \leq 10$. 

Fig.~\ref{fig:mach_cool} (top panel) shows the same as Fig.~\ref{fig:sf} (top panel), but for simulations with different Mach number and for the cooling simulations, at different numerical resolutions. The solid line shows Eq.~(\ref{eq:ls}) with $p=1/2$ and $\phis=0.42^{+0.12}_{-0.09}$ as measured from our simulation at $\mach=4.1$ and resolution of $10048^3$ grid cells (black), where the dotted lines show the upper and lower limit of $\phis$. We see that the $\mach=2$ simulations at resolutions of $1152^3$ and $2304^3$ grid cells agree very well with $\phis\sim0.4$ (see Source Data for Fig.~\ref{fig:mach_cool}). The $\mach=8$ simulations also agree with $\phis\sim0.4$ within the error bars, but the sonic scale is not fully resolved in that case, because the sonic scale occurs on relatively small scales (here $\ls/L\lesssim0.005$ for $\mach\sim8$), posing challenges when the resolution is $<10000^3$ grid cells. Thus, at $\mach=8$, we estimate $\phis\sim0.3$ when converged, $\sim30\%$ smaller than in the $\mach=4$ and~2 simulations. The two cooling simulations show a more significant difference, with $\phis$ possibly as low as $0.2$ when fully converged. This would suggest that the sonic scale occurs on a somewhat smaller scale (by $\sim$~factor~2) when cooling is used instead of the isothermal approximation.

This parameter study suggests that our measurement of $\phis\sim0.4$ holds well, if $L$ is assumed to be the driving scale of the turbulence, and for $\mach\lesssim5$, but $\phis$ may be smaller by as much as a factor of 2 for higher Mach numbers and/or if non-isothermal (cooling) gas is considered. Overall, we conclude that Eq.~(\ref{eq:ls}) is a very good model for the sonic scale, with a nearly constant $\phis\sim0.4$ for $\mach\lesssim5$. For $\mach\gtrsim5$, $\phis$ may have some dependence on $\mach$ and on the details of the thermodynamics (such as cooling and chemistry). These regimes deserve further exploration.

Finally, we emphasise that the measurements of $\ls$ discussed here are all done with respect to the \emph{driving scale} ($L$) in Eq.~(\ref{eq:ls}), which is not necessarily the same as the \emph{molecular cloud scale}. If the main driving of turbulence occurs on scales larger than the cloud scale, then there may not be a distinct feature in the velocity dispersion on the cloud scale. As discussed in the main part of the article, we would then find $\phis\sim0.9$, which implicitly assumes that the power law describing the supersonic cascade runs through the cloud scale without any flattening or feature on that scale. However, we would still expect a change in the Mach number on the cloud scale, because the gas turns from $\gtrsim100\,\mathrm{K}$ in the cold neutral medium to $\sim10\,\mathrm{K}$ in the molecular phase, and this will cause a reduction in the sound speed by a factor of $\gtrsim3$, and hence an increase in the Mach number from transsonic (\mbox{$\mach\sim1$--$2$}) [ref.\cite{VazquezSemadeniEtAl2006,HeitschEtAl2008,BanerjeeEtAl2009,AuditHennebelle2010,ZamoraAvilesEtAl2018,PadoanEtAl2016,PanEtAl2016,KoertgenFederrathBanerjee2017,KoertgenFederrathBanerjee2019,MandalFederrathKoertgen2020}] above the cloud scale, to supersonic speeds ($\mach \gtrsim 5$) below the cloud scale\cite{SchneiderEtAl2013}, depending on the exact velocity dispersion and size of the cloud\cite{Larson1981,SolomonEtAl1987,HeyerBrunt2004,RomanDuvalEtAl2011}. Thus, while the velocity dispersion -- size relation may not show a characteristic change on molecular cloud scale, the Mach number would, and for that reason, $\phis$ is expected to be less than unity, even if we replace the driving scale $L$ in Eq.~(\ref{eq:ls}) with the molecular cloud scale. However, this would need to be investigated in more detail by dedicated simulations and observations, in order to construct a Mach number -- size relation, in addition to the already existing velocity dispersion -- size relation.

\end{methods}

\section*{Supplementary Information}

\section*{The distribution of interstellar filaments}
\begin{figure}
\centerline{\includegraphics[width=0.75\linewidth]{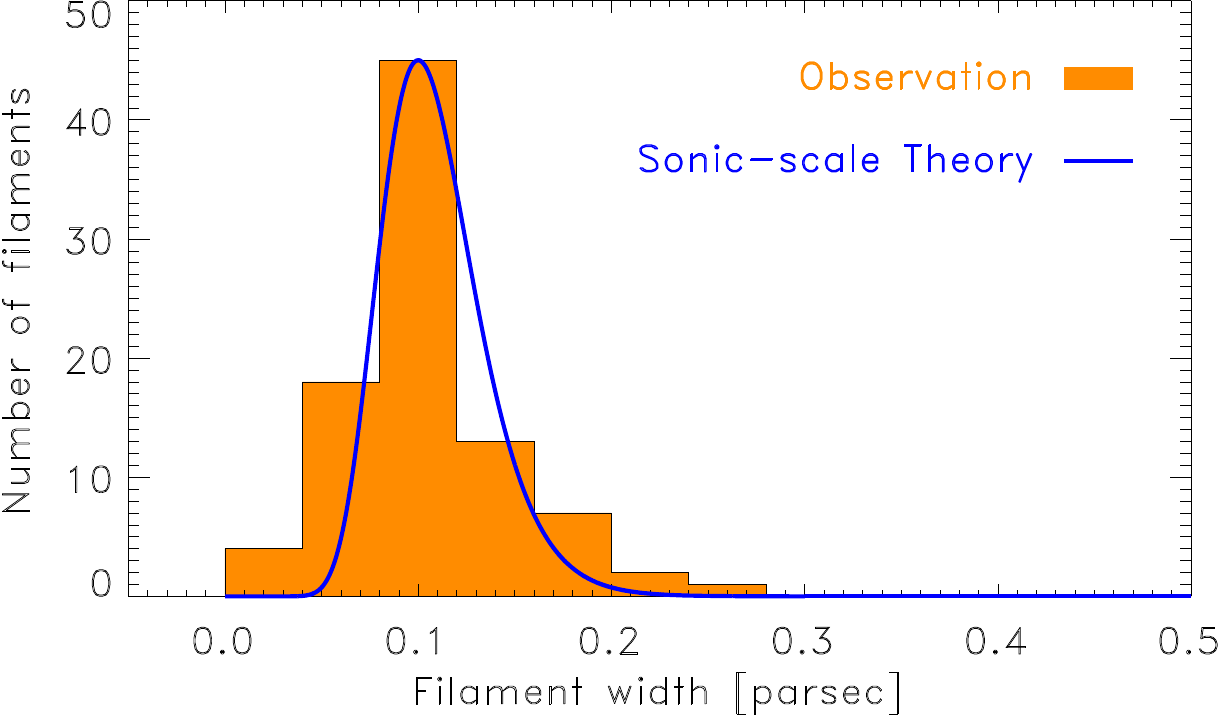}}
\caption{\label{fig:filamentpdf}
Filament width distribution measured from observations in the IC5146, the Aquila and Polaris molecular clouds\protect\cite{ArzoumanianEtAl2011} shown as the orange histogram, together with the theoretical prediction based on the sonic scale (blue line; see {\suppl} for the details of this function). The theoretical prediction is not a fit; instead the peak position is set to $x_0=0.1\,\pc$ and the log-normal standard deviation, $\sigma_x = 0.105$, is determined by the width of the sonic-scale transition measured in Fig.~\ref{fig:sf}.}
\end{figure}

\vspace{-0.5cm}
Our measurement of $\ls$ in the main part of the Letter provides a theoretical prediction of the filament width distribution in molecular clouds. Filaments are the building blocks of molecular clouds\cite{AndreEtAl2010}. Star clusters form preferentially at the junctions of these filaments\cite{SchneiderEtAl2012}. Providing a theoretical model of filament properties is therefore important to further our understanding of the structure of the molecular phase of the interstellar medium. The filament width distribution has been measured in observations of the IC5146, the Aquila and Polaris molecular clouds\cite{ArzoumanianEtAl2011}, and is shown in orange in Fig.~\ref{fig:filamentpdf}. The peak position of the distribution is at a filament width of $\sim0.1\,\pc$. This is coincident with the sonic scale of the clouds\cite{Federrath2016}. However, so far the exact shape and standard deviation of the filament distribution is unconstrained by theory. In main part of the Letter we determined the width of $\ls$, which can be directly translated to the standard deviation of the filament distribution,
\begin{equation} \label{eq:filamentpdf}
N=N_0 \frac{1}{\sqrt{2\pi\sigma_x^2}} \exp\left( -\frac{(\log_{10}x/x_0)^2}{2\sigma_x^2}\right),
\end{equation}
where $N_0$ is the normalisation determined by the number of filaments observed, the characteristic filament width $x_0=0.1\,\pc$, and the log-normal standard deviation of the sonic-scale transition measured in the main part of the Letter, $\sigma_x = \log_{10}(3/2.355)=0.105$. Eq.~(\ref{eq:filamentpdf}) is shown as the blue line in Fig.~\ref{fig:filamentpdf} and provides a good match to the observed filament width distribution. Note, however, that in addition to $\sigma_x$, solely based on the extent of the sonic transition determined here, there may be additional broadening of the filament widths distribution, because of observational uncertainties. Concerning the choice of $x_0=0.1\,\pc$, we note that simulations `\mbox{Cooling $1152^3$}' and `\mbox{Cooling $2304^3\,(1)$}' in Fig.~\ref{fig:mach_cool} produce a sonic scale of \mbox{$\ls\sim0.1\,\pc$} and $0.09\,\pc$, respectively, as a natural consequence of driving turbulence in a multi-phase interstellar medium with cooling (for further details, see Methods in the main part of the Letter).

\section*{The critical density for star formation}
\vspace{-0.5cm}
The sonic scale may also be a key ingredient for predicting the birth rate of stars. It is believed that $\ls$ marks the transition from turbulence-dominated to gravity-dominated clouds, so-called `dense cores', which can collapse to form stars\cite{VazquezBallesterosKlessen2003}. In the model by Krumholz \& McKee\cite{KrumholzMcKee2005}, the critical density for star formation is defined where $\ls$ equals the Jeans scale \mbox{$\lj=[\pi\cs^2/(G\rho_0)]^{1/2}$}, with the cloud mean density $\rho_0$ and the gravitational constant $G$, i.e., $\lj$ is the scale on which a cloud can collapse under the action of its self-gravity\cite{Jeans1902}. Krumholz \& McKee assumed that $\ls$ is given by $\ls^\mathrm{KM}=L\mach^{-2},$ i.e., $\phis=1$ in Eq.~1 from the main part of the Letter. This led to the definition of the critical density\cite{KrumholzMcKee2005,FederrathKlessen2012,KainulainenFederrathHenning2014},
\begin{equation} \label{eq:rhocrit}
\rhocrit = \left(\phix\frac{\lj}{\ls^\mathrm{KM}}\right)^2 \rho_0 = (\phix\phis)^2 \left(\frac{\lj}{\ls}\right)^2 \rho_0,
\end{equation}
where we have inserted Eq.~1 from the main part of the Letter, i.e., the true $\ls=\phis\ls^\mathrm{KM}$ in the second equality. The `fudge factor' $\phix=0.18\pm0.03$ was determined based on fitting a comprehensive parameter study of star-formation simulations\cite{FederrathKlessen2012}. Together with our measurement of $\phis=0.42^{+0.12}_{-0.09}$ in the main part of the Letter, this shows that the critical density for star formation occurs where $\lj/\ls = (\phix\phis)^{-1} = 13^{+7}_{-4}$, if $L$ is assumed to be both the driving scale of the turbulence and the molecular cloud scale. If the driving scale is on scales larger than the cloud scale, and the supersonic power-law cascade extends through the cloud scale $L$, we have $\phis=0.91^{+0.25}_{-0.20}$ (see main part of the Letter), and we find $\lj/\ls = (\phix\phis)^{-1} = 6.1^{+3.2}_{-2.0}$. Thus, in both cases, $\rhocrit$ is on a somewhat larger scale and in somewhat lower-density gas than defined by the assumption $\ls = \lj$, i.e., on scales up to about an order of magnitude larger than $\ls$. This implies that low-density, large-scale gas flows on scales about an order of magnitude larger than the sonic scale may significantly contribute to star formation -- and not only gas below the sonic scale.

\begin{addendum}

\item[Acknowledgements:] C.F.~acknowledges funding provided by the Australian Research Council (Discovery Project DP170100603 and Future Fellowship FT180100495), and the Australia-Germany Joint Research Cooperation Scheme (UA-DAAD). C.F.~further acknowledges the Australian Research Council Centre of Excellence for All Sky Astrophysics in 3~Dimensions (ASTRO~3D), through project number CE170100013.
R.S.K.~acknowledges support from the German Research Foundation (DFG) via the collaborative research center ``The Milky Way System'' (SFB~881, Project-ID~138713538, subprojects A1, B1, B2, and B8) as well as support from the Heidelberg cluster of excellence EXC~2181 (Project-ID~390900948) ``STRUCTURES: A unifying approach to emergent phenomena in the physical world, mathematics, and complex data'' funded by the German Excellence Strategy. He also thanks the European Research Council for support via the ERC Advanced Grant ``STARLIGHT'' (project ID~339177) and the ERC Synergy Grant ``ECOGAL'' (project ID~855130). We further acknowledge high-performance computing resources provided by the Leibniz Rechenzentrum and the Gauss Centre for Supercomputing (grants~pr32lo, pr48pi and GCS Large-scale project~10391), the Australian National Computational Infrastructure (grant~ek9) in the framework of the National Computational Merit Allocation Scheme and the ANU Merit Allocation Scheme. The simulation software FLASH was in part developed by the DOE-supported Flash Center for Computational Science at the University of Chicago.

\end{addendum}

\section*{References}
    
\def\apj{{\sl Astrophys.~J.}}
\def\apjl{{\sl Astrophys.~J.}}
\def\apjs{{\sl Astrophys.~J.~Suppl.~Ser.}}
\def\aap{{\sl Astron.~Astrophys.}}
\def\aaps{{\sl Astron.~Astrophys.~Suppl.~Ser.}}
\def\aj{{\sl Astron.~J.}}
\def\araa{{\sl Annu.~Rev.~Astron.~Astrophys.}}
\def\mnras{{\sl Mon.~Not.~R.~Astron.~Soc.}}
\def\physrep{{\sl Phys.~Rep.}}
\def\prl{{\sl Phys.~Rev.~Lett.}}
\def\jfm{{\sl J.~Fluid Mech.}}
\def\rmp{{\sl Rev.~Mod.~Phys.}}
\def\jcp{{\sl J.~Comput.~Phys.}}
\def\aapr{{\sl Astron.~Astrophys.~Rev.}}
\def\pasj{{Publ.~Astron.~Soc.~Japan}}


\end{document}